\documentclass[english,aps,prl,showpacs,twocolumn,superscriptaddress,preprintnumbers,nofootinbib]{revtex4-1}

\global\long\def\order#1{\mathcal{O}\left(#1\right)}

\usepackage{amsmath}
\usepackage{amssymb}

\usepackage{slashed}
\usepackage{graphicx}
\usepackage{epsfig}

\usepackage{epstopdf}
\usepackage{esint}

\usepackage{bm}
\usepackage{scalerel}
\def\subH{\scaleobj{.8}{\text{H}}}
\def\subEH{\scaleobj{.8}{\text{EH}}}

\begin{document}

\title{Logarithmically enhanced Euler--Heisenberg Lagrangian
  contribution to the electron gyromagnetic factor} 

\author{Andrzej Czarnecki}
\affiliation{Department of Physics, University of Alberta, Edmonton, Alberta, Canada T6G 2E1}

\author{Jan Piclum}
\affiliation{Theoretische Physik 1, Naturwissenschaftlich-Technische Fakult\"at, Universit\"at Siegen, 57068 Siegen, Germany}

\author{Robert Szafron}
\email{robert.szafron@cern.ch}
\affiliation{Theoretical Physics Department, CERN, 1211 Geneva 23, Switzerland}

\begin{abstract}
  Contrary to what was previously believed, two-loop radiative corrections to the
  $g$-factor of an electron bound in a hydrogen-like ion at
  $\mathcal{O}\left(\alpha^2 (Z\alpha)^5\right)$ exhibit logarithmic
  enhancement. This previously unknown contribution is due to a
  long-distance light-by-light scattering amplitude. Taking an
  effective field theory approach, and using the Euler--Heisenberg Lagrangian, we find
  $\Delta g =  -\left( \frac{\alpha}{\pi} \right)^2
  \left(Z\alpha\right)^5 \frac{56 \pi}{135}\ln Z\alpha$.
\end{abstract}

\preprint{Alberta Thy 2-20, SI-HEP-2020-21, P3H-20-042, CERN-TH-2020-137}


\maketitle


The gyromagnetic factor $g$ describes the proportionality of a
particle's magnetic moment $\bm \mu$ to its spin $\bm s$,
\begin{equation}
  \label{eq:1}
  \bm \mu = g {e \over 2m} \bm s,
\end{equation}
where $e$ is the particle's charge, $m$ its mass, and we use
units $c=\hbar  = 1$. Boldface letters denote usual vectors.

We consider the electron, for which 
Dirac's theory predicts $g=2$. This value is corrected at the per
mille level by the electron's
self-interactions, at present known to the fifth order in the fine
structure constant $\alpha \simeq 1/137$ \cite{Aoyama:2019ryr,Laporta:2017okg}.

Larger corrections can arise from the electron's interaction with its
environment. The simplest such influence is the Coulomb field of the
nucleus to which the electron is bound.
These binding corrections are reviewed in
Ref.~\cite{Zatorski:2017vro}. 
Full numerical evaluation of two-loop self-energy diagrams is under way 
\cite{Sikora:2018zda} and first results for a class of diagrams are 
already available \cite{Yerokhin:2013qma,Debierre:2020zey}. 

These calculations are of
great metrological interest, because the electron mass ($m$ in Eq.~\eqref{eq:1}) is best
determined with an ion in a Penning trap, rather than by trapping an
electron alone (binding to a nucleus greatly decreases errors caused
by the electron's thermal motion). In the future, a competitive
value of the fine structure constant may also be obtained from such
measurements \cite{Sturm:2019fuw,shabaev2006g,Yerokhin:2016gxj},
complementary to atom interference \cite{Parker2018} and the
free-electron $g-2$ \cite{Gabrielse:2019cgf}. The potential of the bound
$g$-factor to constrain scenarios beyond the Standard Model is
discussed in Ref.~\cite{Debierre:2020wky}.

Neglecting nuclear structure corrections, $g$ can be expressed in a double
series in powers of $\alpha/\pi$ (self-interactions) and in powers and
logarithms of  $Z\alpha$ (interactions with the nucleus). With $L=-\ln( Z\alpha)^2$,
\begin{align}
  \label{eq:7}
  g& = {2\over 3}\left[ 1 + 2\sqrt{1-(Z\alpha)^2} \right]
+\frac{\alpha}{\pi}\sum_{i,j=0}^\infty  a_{ij} \left(Z\alpha\right)^i
     L^j 
\nonumber \\
&+\left(\frac{\alpha}{\pi}\right)^2 \sum_{i,j=0}^\infty  b_{ij}
               \left(Z\alpha\right)^i L^j 
+ \order{ \left(\frac{\alpha}{\pi}\right)^3 }.
\end{align}
This structure mirrors the expansion of atomic energy levels (Lamb
shift \cite{Yerokhin:2018gna}) and so far it has been found that if a logarithm is present in one
observable in a given order, it is also present in the other.  This
rule is very important because the Lamb shift is better
understood theoretically than the $g$ factor. Measurements with ions
of various $Z$ have been used to fit unknown coefficients in
Eq.~\eqref{eq:7} \cite{Zatorski:2017vro} to extract the electron mass.

Here we find the first exception from this rule: in the Lamb shift the
coefficient corresponding to 
$b_{51}$ vanishes, whereas we find that
\begin{equation}
  \label{eq:9}
  b_{51} = {28\pi\over 135}.
\end{equation}
In principle, logarithmic effects can always be calculated in at least
two ways. The argument $Z\alpha$ is really a ratio of two distance
scales, for example, the large Bohr radius and the small electron Compton wavelength.
One can calculate only the long-distance or short-distance
part. In both cases one finds the same magnitude of logarithmic
divergence. 

In the present case, we did both, to be sure that the
logarithmic contribution really exists. Below we briefly outline both
parts of the calculation.
We leave for the future work the evaluation of the non-logarithmic 
part, together with providing further technical details of the 
computation.

\begin {figure}[htb]
\includegraphics[width=0.48\columnwidth]{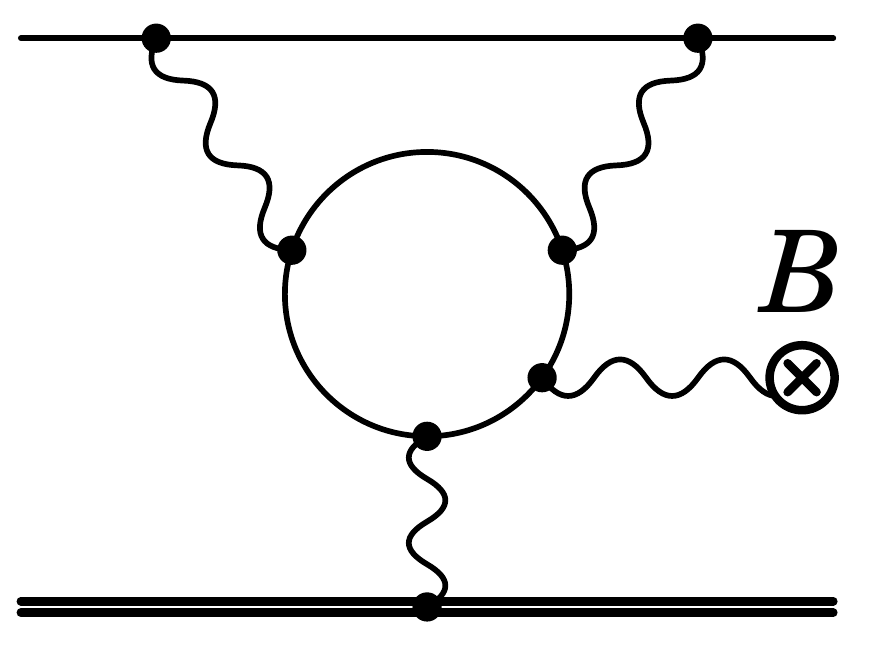}
\caption{\label{fig:diagram1} An LBL loop contributing to $\alpha^2$
  corrections to the bound electron $g$-factor. The thick double line
  denotes the nucleus. Thin solid lines are electrons and wavy lines
  are photons, including the external magnetic field $\bm B$.}
\end{figure}

The new logarithmic contribution is an effect of the virtual
light-by-light scattering (LBL). It arises through the coupling of
four photons induced by their interaction with a virtual charged
particle such as an electron. Figure \ref{fig:diagram1} provides an
example, where the Coulomb field of the nucleus couples to an external
magnetic field, and two photons are interacting with the bound electron,
thus changing the electron's $g$-factor and modifying its response to
the magnetic field.

LBL was first predicted by Heisenberg and Euler
\cite{Heisenberg:1935qt,Euler:1935zz} and by Weisskopf
\cite{Weisskopf:1996bu} who determined corrections to Maxwell's
Lagrangian of the electromagnetic field $\mathcal {L} = {1\over 2}\left({\bm E}^{2}-{\bm B}^{2}\right)$,
\begin{align}
\mathcal{L}_{\subEH} =  \frac{\alpha^{2}}{m^{4}}
\left[c_{1}\left({\bm E}^{2}-{\bm  B}^{2}\right)^{2}
+c_{2}\left({\bm E}\cdot{\bm B}\right)^{2}\right], 
\label{EH}
\end{align}
with $c_1=\frac{2}{45}$ and $c_2=\frac{14}{45}$. While $\mathcal {L}$
leads to linear Maxwell's equations, $\mathcal{L}_{\subEH}$
introduces non-linear effects. Nowadays, this classic result is often the first non-trivial example encountered by students learning effective field theory methods. 

Searches for effects of $\mathcal{L}_{\subEH} $ have so far been in
vain \cite{Ejlli:2020yhk}. 
Observed  non-linear effects arise either from
interactions with matter (non-linear optics) or from high-energy
processes with photon momenta much larger than the electron mass,
beyond the validity of $\mathcal{L}_{\subEH}$. For
example, photon splitting $\gamma N\to\gamma\gamma N$ has been
measured \cite{Akhmadaliev:2001ik} (see \cite{Lee:2001gc} for a
theoretical review). A related process is Delbr\"uck scattering $\gamma
N\to\gamma N$ \cite{Milstein:1994zz}. The high-energy LBL scattering has been observed in ultraperipheral heavy-ion collisions \cite{Aad:2019ock}. 

The $\mathcal{L}_{\subEH}  $ effect described in this paper likely has the best chance of being
experimentally accessible.


It often happens with bound state radiative corrections that a single
Feynman diagram contributes to different orders in the perturbative
expansion. To disentangle corrections of different orders, it is
convenient to use the expansion by regions
\cite{Tkachov:1997gz,Beneke:1997zp,Smirnov:2002pj,Czarnecki:1996nr,Czarnecki:1997vz}. 
Once the relevant modes are identified, a
systematic expansion can be achieved by setting up an effective field
theory (EFT) whose operators capture the low-energy physics, while the
so-called matching coefficients contain information about 
short-distance phenomena. 

In bound-state quantum electrodynamics (QED), the relevant EFT is
obtained in a two-step process. First, we integrate out the hard
modes, i.e.~momenta of the order of electron mass $m$. The resulting
theory is known as non-relativistic QED (NRQED), introduced by Caswell
and Lepage \cite{Caswell:1985ui}. The NRQED Lagrangian is organised in
powers of the electron's velocity (in an ion with the atomic number $Z$,
that velocity is $v\sim Z\alpha$), or inverse powers of electron mass
\cite{Hill:2012rh}. The Euler–Heisenberg (E-H) Lagrangian
$\mathcal{L}_{\subEH}$ is part of the NRQED
Lagrangian and contributes at $\mathcal{O}(v^4)$.

NRQED is still complicated and contains modes with a range of energy
scales. In the second step, one  integrates out soft modes whose
momenta scale as $mv$, and potential photons with energy
$E\sim m v^2$ and three-momentum ${\bm p}\sim m v $. The resulting
theory is called potential-NRQED (PNRQED)
\cite{Pineda:1997ie,Pineda:1998kn,Beneke:1999zr}.  It contains
instantaneous, non-local interactions between the electron and the
nucleus, the so-called potentials.

The leading one is the Coulomb potential responsible
for the binding and described by the operator 
\begin{align}\label{eq:pot_c}
\int d^3 r \left[\chi^\dagger_e \chi_e\right]({\bm x}+{\bm r})
\left(- \frac{Z\alpha}{r}\right)\left[N^\dagger N\right]({\bm x}),
\end{align} 
where $\chi_e$ is the non-relativistic electron field, and $N$ is the nucleus field.
Other potentials are treated as perturbations. 

To compute the contribution of the E-H interaction to the bound
electron $g$-factor, we have to generalise potentials to include
spin-dependent interactions with an external magnetic field.  The
two-step EFT approach has been successfully used to compute
spin-independent observables before. Technical details can be found in
Ref.~\cite{Szafron:2019tho} (see also
\cite{Jentschura:2005xu,Pachucki:2004si,Pachucki:2005px,Pachucki:2004zz,Peset:2015zga,Peset:2017ymh}).

LBL scattering first contributes to the
bound electron $g$-factor at
$\mathcal{O}\left(\alpha (Z\alpha)^5 \right)$
\cite{Karshenboim:2002jc} and
$\mathcal{O}\left(\alpha^2 (Z\alpha)^4\right)$
\cite{Czarnecki:2016lzl}. In both these cases, the LBL scattering was
a part of a short-distance correction to the bound electron
$g$-factor. Here we focus on the former type of diagrams, where two
photons are attached to the electron line.

\begin{figure}[htb]
\includegraphics[width=0.45\columnwidth]{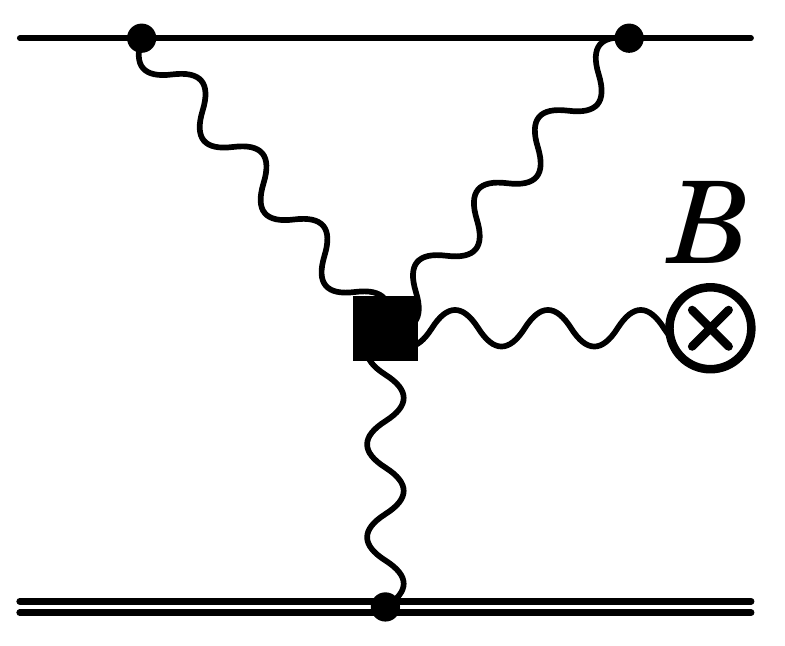}
\caption{\label{fig:diagNRQED} NRQED diagram corresponding to Figure
  \ref{fig:diagram1} with the Euler--Heisenberg Lagrangian insertion
  replacing the electron loop. }
\end{figure}

We start by analysing the diagram in Figure \ref{fig:diagram1}.  The
case where both loops are hard was discussed in
\cite{Czarnecki:2016lzl}. In that  case, the loops collapse to a
point in NRQED, where the diagram is represented by an effective
operator with two photon fields. This operator is then matched on the
effective spin-dependent potential.  Here we consider a
situation where only the fermionic loop is hard, while the second loop
is soft. This means that only the LBL fermionic loop is a short-distance phenomenon, while photons are part of the long-distance physics.  The hard matching leads to the E-H Lagrangian in Eq.~\eqref{EH}.
The soft loop in the QED diagram is now represented in NRQED by a time-ordered
product of the E-H Lagrangian  
\begin{align}\label{eq:Tprod}
\frac{1}{3!} \int d^4x \int d^4 y \int d^4 z T\left[\mathcal{L}_{\text{I}}(x) ,\mathcal{L}_{\text{I}}(y),  \mathcal{L}_{\subEH}(z)  \right],
\end{align}
with the interaction Lagrangian containing the leading Coulomb interaction and a Pauli interaction
\begin{align}
\mathcal{L}_{\text{I}} 
= \chi_e^{\dagger}\left(-e A_0+c_{F}e\frac{{\bm \sigma}\cdot{\bm B}}{2m}\right)\chi_e.
\end{align}
Here $c_F=1+\frac{\alpha}{2 \pi}+\mathcal{O}(\alpha^2)$ and
${\bm \sigma}$ is a vector composed of Pauli matrices.  The NRQED
diagram representing the time-ordered product (\ref{eq:Tprod}) is
depicted in Figure \ref{fig:diagNRQED}.

To perform the second matching step, we compute the amplitude for the diagram shown in Figure~\ref{fig:diagNRQED}; it reads 
\begin{align}
ic_2 c_F\frac{\alpha^{2}}{m^{4}}\frac{Ze^{3}}{32m}\frac{Q_{i}Q_{j}}{\left|{\bm Q}\right|}\hat{\chi}_e^{\dagger}\sigma_j\hat{\chi}_eB_i,
\end{align}
with ${\bm Q}$ representing the momentum transfer between the electron and the
nucleus, and the external magnetic field ${\bm B}$ that
carries zero momentum. $\hat{\chi}_e$ denotes non-relativistic electron spinors. 

We drop the part of the amplitude that does not contribute in
$s$-states and, after a Fourier transform, we find the spin-dependent
correction to the PNRQED potential,
\begin{align}
e \int d^3 r \left[\chi^\dagger_e 
\frac{{\bm \sigma}\cdot {\bm  B}_{\text{ext}}}{2m} 
\chi_e\right]({\bm x}+{\bm r})\,\delta V(r)\left[N^\dagger N\right]({\bm x}),
\end{align} 
with
\begin{align}
\delta V\left(r\right)= - c_{2}c_F\frac{\alpha^{2}}{\pi^{2}}
\frac{Z\alpha}{\left(mr\right)^{4}}\frac{\pi}{12}.
\end{align}
This potential has $r^{-4}$ dependence and it is thus more singular
for small $r$ than the leading Coulomb potential in
Eq.~(\ref{eq:pot_c}). Consequently, the matrix element in an $s$ state
is divergent and has to be regularised. We use dimensional
regularisation with space-time dimension $D=4-2\epsilon$ and find the
E-H contribution to the bound electron $g$-factor to be
\begin{align}\label{eq:gEH}
\Delta g_{\subEH}=\left(\frac{\alpha}{\pi}\right)^{2}
\frac{28 \pi }{135}
\left(Z\alpha\right)^{5}\left( {1 \over\epsilon}-
\ln\frac{\left(mZ\alpha\right)^{2}}{\mu^2}+\ldots\right),
\end{align}
where dots represent terms that are not logarithmically enhanced. The
computation of the matrix element is closely related to the
logarithmic correction to the Lamb shift described in
\cite{Czarnecki:2016lzl}.  

\begin {figure}[htb]
\includegraphics[width=0.48\columnwidth]{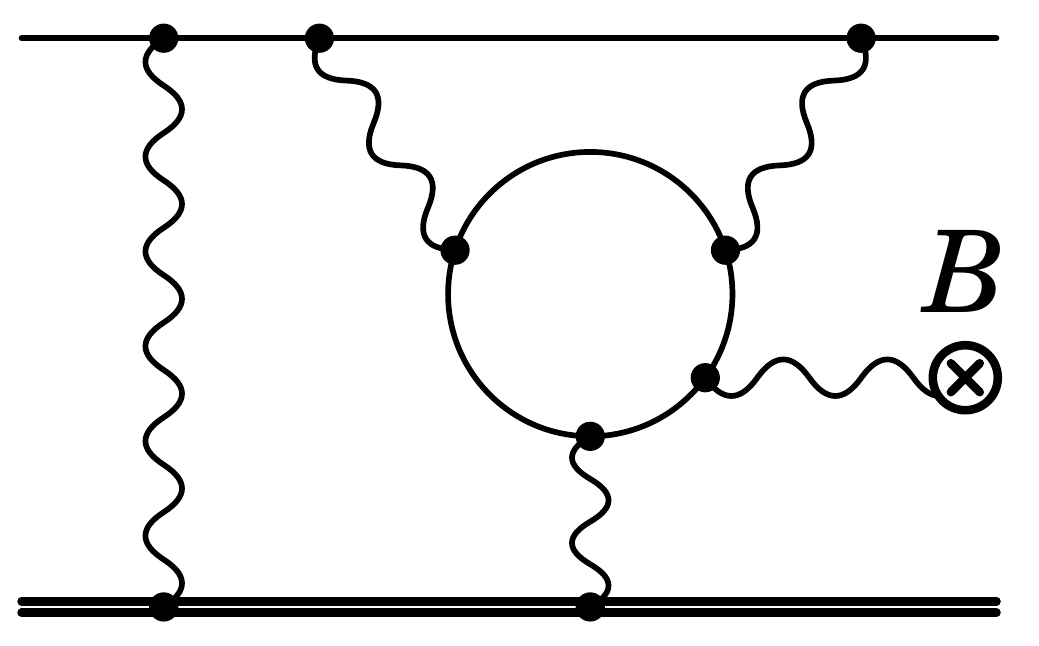}\hspace*{5mm}
\includegraphics[width=0.48\columnwidth]{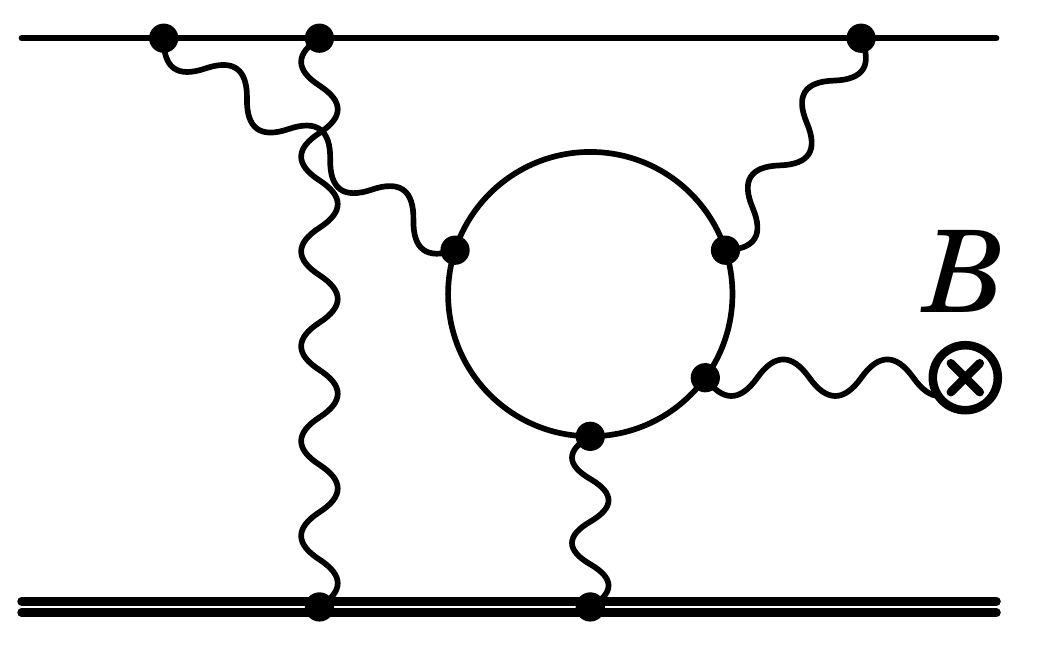}
\caption{\label{fig:diagram2} The high-energy correction to the diagram
in Figure  \ref{fig:diagram1} arises from an additional hard photon exchanged
  between the electron and the nucleus. Two examples are shown. Other
  diagrams are found by permuting photons coupled to each electron line. The diagram on the right does not contribute to the divergent part in Eq. (\ref{eq:gH}).}. 
\end{figure}

The $1/\epsilon$ ultraviolet (UV) pole of the matrix element of $\delta V$
cancels with the high-energy contribution shown in
Figure \ref{fig:diagram2}. The additional photon connecting the external
electron to the nucleus may be understood as a high-energy
tail of the electron wave function. This is why only the diagrams related to the left diagram in Figure \ref{fig:diagram2} contribute to the divergent part. 
In this short distance part of the
correction all loop momenta  have a
hard scaling ($\sim m$). Ref.~\cite{Pachucki:2017xfd} explains the theory of
the high-energy contribution to the bound $g$-factor at
$\mathcal{O}((Z\alpha)^5)$.

In the short distance calculation we proceed as in our
previous calculations~\cite{Dowling:2009md,Czarnecki:2017kva}. All
three-loop integrals are reduced to a small set of master integrals
with the so-called Laporta
algorithm~\cite{Laporta:1996mq,Laporta:2001dd} implemented in the
program {\tt FIRE}~\cite{Smirnov:2014hma}. Even though we are dealing
with diagrams that do not contribute to the Lamb shift, almost all
master integrals are the same as before, and their results can be found
in~\cite{Dowling:2009md}. The reason is that master integrals
correspond to scalar diagrams, where some of the lines are absent. In
most cases, one can transform these master integrals into 
known ones. 

However, there is one new master integral,
\begin{align}
  \label{eq:5}
& 
  \int \frac{\mathrm{d}^Dk_1\, \mathrm{d}^Dk_2\, \mathrm{d}^Dk_3
  \;  \delta(k_2^0)}{ k_1^2\, (k_1-k_2)^2\, (k_3^2+m^2)\,
             [(k_2+k_3)^2+m^2] }
\nonumber\\ 
 &
= -  \frac{64\pi^7m^3}{3} + {\mathcal O}(\epsilon), 
\end{align}
that could not be checked with previous calculations. For this reason the
computation of the hard part alone would not be a sufficient proof of
the presence of the logarithm. Fortunately, the hard correction we found,
\begin{align}\label{eq:gH}
\Delta g_{\subH}
=   -\left(\frac{\alpha}{\pi}\right)^{2}\frac{28}{135}\pi
  \left(Z\alpha\right)^{5}\left( {1\over\epsilon} - \ln\frac{m^{2}}{\mu^2}+\ldots \right).
\end{align}
is consistent with the soft correction in Eq.~(\ref{eq:gEH}). Summing Eqs. (\ref{eq:gEH}) and (\ref{eq:gH}) we find that $1/\epsilon$
singularities cancel and obtain our main result,
\begin{align}
\Delta g(Z) = \Delta g_{\subH}+\Delta g_{\subEH}
=
  \left(\frac{\alpha}{\pi}\right)^{2}\left(Z\alpha\right)^{5}\frac{28}{135}\pi\ln{1\over
  (Z\alpha)^2} \, ,
\end{align}
from which we read off the coefficient $b_{51}$ in Eq.~\eqref{eq:9}.


Due to the logarithmic enhancement, the correction is much larger than
anticipated and exceeds other LBL corrections computed previously in
\cite{Czarnecki:2017kva}. 

For the hydrogen-like carbon ion, currently the best source of the
electron mass determination, the resulting relative correction to the
$g$-factor and, by the same token, to the electron mass $m$, is
\begin{equation}
  \label{eq:2}
  {\Delta g(Z = 6) \over g} = {\Delta m \over m} = 1.8\cdot 10^{-12},
\end{equation} 
about 17 times smaller than the current experimental
error. This correction will likely become important for the
measurements in the near future \cite{Sturm:2019fuw}.

Because of the factor $Z^5$, the correction grows rapidly for heavier
ions. For the experimentally important silicon
\cite{PhysRevLett.107.023002},
\begin{equation}
  \label{eq:3}
   {\Delta g(Z = 14) \over g} = 0.9\cdot 10^{-10},
\end{equation}
exceeding the accepted theoretical uncertainty of $0.7\cdot 10^{-10}$
\cite{Zatorski:2017vro}. 
This is likely because the Ref.~\cite{Zatorski:2017vro} fitted unknown
higher-order corrections, assuming a vanishing $b_{51}$, as we
explained below Eq.~\eqref{eq:7}.

For the future, two extensions of this work are of interest. While we
have determined the E-H effect in a one-electron hydrogen-like ion,
few-electron systems, especially lithium- and boron-like ions, are
also experimentally relevant \cite{Glazov:2019utw}.
It would also be interesting to evaluate the E-H correction for a
muonic atom \cite{Sikora:2018laj} where it should be
further enhanced by the logarithm of the electron to
muon mass ratio.

\section*{Acknowledgments}
RS would like to thank Martin Beneke for useful discussions. The loop
diagrams were calculated with FORM \cite{Ruijl:2017dtg,Vermaseren:2000nd}.   This
work was supported by the Natural Sciences and Engineering Research
Council of Canada and by the Deutsche Forschungsgemeinschaft (DFG, German Research Foundation) under grant  396021762 - TRR 257.

\end{document}